%% file: jsc.tex
\documentclass[sn-mathphys-num]{sn-jnl}

\usepackage{amsmath,amssymb,amsfonts}
\usepackage{graphicx}
\usepackage{textcomp}
\usepackage{xcolor}
\usepackage{algorithm}
\usepackage[noend]{algpseudocode}
\usepackage{graphicx}
\usepackage{textcomp}
\usepackage{xcolor}
\usepackage{xparse}
\usepackage{url}
\usepackage{siunitx}
\usepackage{fancyvrb}   
\usepackage{listings}
\usepackage{subcaption}
\usepackage{todonotes}
\usepackage{enumitem}
\usepackage{booktabs}
\usepackage{lscape}

\definecolor{codegreen}{rgb}{0,0.6,0}
\definecolor{codegray}{rgb}{0.5,0.5,0.5}
\definecolor{codepurple}{rgb}{0.58,0,0.82}
\definecolor{backcolour}{rgb}{0.95,0.95,0.92}
\lstdefinestyle{mystyle}{
    backgroundcolor=\color{backcolour},   
    commentstyle=\color{codegreen},
    keywordstyle=\color{magenta},
    numberstyle=\tiny\color{codegray},
    stringstyle=\color{codepurple},
    basicstyle=\ttfamily\small,
    breakatwhitespace=false,         
    breaklines=true,                 
    captionpos=b,                    
    keepspaces=true,                 
    numbers=left,                    
    numbersep=5pt,                  
    showspaces=false,                
    showstringspaces=false,
    showtabs=false,                  
    tabsize=2,
    frame=single,
    xleftmargin=10pt,
    xrightmargin=7pt,
    morekeywords={write, fsync}
}
\begin{document}

\title{Recorder: Comprehensive Parallel I/O Tracing and Analysis}

\author*[1]{\fnm{Chen} \sur{Wang}}\email{wang116@llnl.gov}

\author[2]{\fnm{Izzet} \sur{Yildirim}}\email{iyildirim@hawk.iit.edu}

\author[1]{\fnm{Hariharan} \sur{Devarajan}}\email{hariharandev1@llnl.gov}

\author[1]{\fnm{Kathryn} \sur{Mohror}}\email{mohror1@llnl.gov}

\author[3]{\fnm{Marc} \sur{Snir}}\email{snir@illinois.edu}

\affil*[1]{\orgdiv{Center for Applied Scientific Computing}, \orgname{Lawrence Livermore National Laboratory}, \orgaddress{\street{7000 E. Ave}, \city{Livermore}, \postcode{94550}, \state{California}, \country{USA}}}

\affil[2]{\orgdiv{Department of Computer Science}, \orgname{Illinois Institute of Technology}, \orgaddress{\street{10 W 35th St}, \city{Chicago}, \postcode{60616}, \state{Illinois}, \country{USA}}}

\affil[3]{\orgdiv{Department of Computer Science}, \orgname{University of Illinois Urbana-Champaign}, \orgaddress{\street{201 N Goodwin Ave}, \city{Urbana}, \postcode{61801}, \state{Illinois}, \country{USA}}}


\input{sections/abstract}


\maketitle

\section{Introduction}\label{sec:introduction}
\input{sections/introduction}

\section{Overview}\label{sec:overview}
\input{sections/overview}

\section{Pattern-Recognition-Based Compression}\label{sec:compression}
\input{sections/compression}

\section{Enabling In-depth Analyses}
\label{sec:enabled-analyses}
\input{sections/enabled_analyses}

\section{Evaluation}\label{sec:evaluation}
\input{sections/evaluation}

\section{Related Work}\label{sec:related-work}
\input{sections/related_work}

\section{Conclusion and Future Work}\label{sec:conclusion}
\input{sections/conclusion}

\section*{Acknowledgment}
\input{sections/acknowledgment}

\bibliography{references}

\end{document}

%% file: sections/abstract.tex
\abstract{
This paper presents Recorder, a parallel I/O tracing tool designed to capture comprehensive I/O information on HPC applications. 
Recorder traces I/O calls across various I/O layers, storing all function parameters for each captured call. The volume of stored information scales linearly the application's execution scale. To address this, we present a sophisticated pattern-recognition-based compression algorithm. This algorithm identifies and compresses recurring I/O patterns both within individual processes and across multiple processes, significantly reducing space and time overheads.
We evaluate the proposed compression algorithm using I/O benchmarks and real-world applications, demonstrating that Recorder can store more information while requiring approximately 12× less storage space compared to its predecessor. Notably, for applications with typical parallel I/O patterns, Recorder achieves a constant trace size regardless of execution scale. Additionally, a comparison with the profiling tool Darshan shows that Recorder captures detailed I/O information without incurring substantial overhead. The richer data collected by Recorder enables new insights and facilitates more in-depth I/O studies, offering valuable contributions to the I/O research community.
\looseness=-1

}

%% file: sections/introduction.tex
\begin{figure}[htbp]
    \centering
    \includegraphics[width=0.5\linewidth]{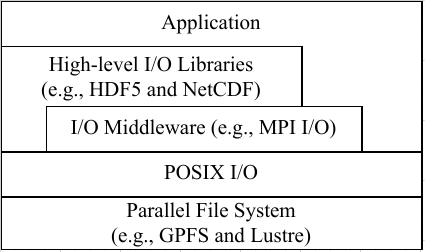}
    \caption{Parallel I/O Software Stack}
    \label{fig:io_stack}
\end{figure}

The I/O subsystem is a critical component of HPC systems.
HPC applications interact with the I/O subsystem through the interface provided by the underlying file system, usually making use of the POSIX~\cite{POSIX} interface. Alternatively, they can opt for high-level I/O libraries, such as HDF5~\cite{HDF5} and NetCDF~\cite{rew:1990:netcdf}, which offer a more user-friendly interface for managing scientific data. These libraries are often built upon I/O middleware components (e.g., MPI-IO~\cite{corbett:1995:mpiio}), which, in turn, are implemented using the underlying file system interface. Figure~\ref{fig:io_stack} shows a common parallel I/O software stack found in current HPC systems. The intricate interactions among these multiple I/O software components pose challenges to understanding the I/O behavior of an HPC application, making it particularly difficult to identify I/O bottlenecks.

To address this challenge, various I/O profiling and tracing tools have been developed. Profiling tools (\cite{Darshan, TAU}) collect more coarse-grained information, thus imposing lower overhead. In contrast, tracing tools (\cite{kim2012iopin, Recorder, Recorder2}) capture more detailed information but also tend to incur a higher overhead. In both categories (profiling and tracing), the total overhead consists of space overhead and time overhead. The space overhead includes the memory footprint of the tool at execution time, as well as the size of the produced trace file. The time overhead results from the extra execution time caused by the use of the tool.

Recorder~\footnote{\url{https://github.com/uiuc-hpc/Recorder}} falls into the category of tracing tools, aiming to bridge the gap between the capabilities of existing I/O tracing tools and the demands of large-scale HPC applications and workflows. One notable analysis that motivated the creation of Recorder involves investigating the data flow of HPC workflows. These workflows may comprise diverse components including MPI/non-MPI programs, single-threaded/multi-threaded programs, and CUDA-enabled programs. To our knowledge, none of the existing tracing tools supports all these components while preserving sufficient information (e.g., call chains, I/O request size, synchronized timestamps, etc.), thus failing to provide a comprehensive view of the data flow within the HPC workflow.

To allow such comprehensive I/O studies, we expect Recorder to be equipped with the following capabilities and features:

\begin{itemize}
    \item Capable of tracing I/O calls from multiple layers while preserving all parameters of each call. Most existing tracing tools either trace a single layer or preserve only a subset of function parameters. Tracing a single layer provides only a partial view of the call chain and cannot pinpoint the root cause of the I/O bottlenecks. Preserving a limited number of parameters may lose valuable information, e.g., the datatype of MPI write calls and the flag of open calls.
    \item Ready for tracing large-scale runs of HPC applications. Existing tools often struggle to collect information from large-scale runs, as their produced trace sizes grow linearly with execution time and scale.
    \item Support for multi-threaded programs to offer insights for HPC applications that engage in asynchronous I/O. This feature is often lacking in existing tracing tools.
    \item Support for non-MPI programs, another easily overlooked feature that impedes a tracing tool from offering a comprehensive view of data flow in HPC workflows, which may include both MPI and non-MPI components.
\end{itemize}

However, the previous version of Recorder~\cite{Recorder2}, referred to as \textit{Recorder-old} throughout this paper, has several key limitations and missing features that prevent in-depth analyses of complex HPC applications and workflows.
First, the tracing code in Recorder-old must be manually written for each targeted function, making it error-prone and difficult to extend. Second, Recorder-old uses a peephole-based compression technique, which is inadequate for capturing recurring I/O patterns. Moreover, the compression is performed only within each process, causing the trace size to scale linearly with the number of processes, which can become unmanageable at larger scales. Third, Recorder-old supports only single-threaded MPI programs, limiting its applicability to various scenarios, including HPC workflows that involve multi-threaded and non-MPI programs. Lastly, Recorder-old stores traces in a custom binary format, which poses challenges for analysis using existing post-processing tools.

In this work, we present a major revision of Recorder that addresses previous limitations while providing all the aforementioned features. Our contributions are summarized as follows:
\begin{itemize}
    \item \textbf{Automatic Tracing Code Generation:} We introduce a new tracing mechanism that employs a three-phase generic tracing wrapper for each function of interest. These tracing wrappers can be generated automatically and compiled as plugins, making it extremely easy to extend support for additional functions and I/O libraries. (Section~\ref{subsec:tracing-code-generation}).
    \item \textbf{Supporting Diverse Programs:} We extend Recorder to trace a wide range of program types, including single-threaded, multi-threaded, MPI, non-MPI programs, and CUDA kernels. Recorder also captures rich metadata, such as thread IDs and function call depth, offering a deeper insight into the traced data. These features enable the tracing and analysis of I/O operations across a broader range of HPC scenarios. (Section~\ref{subsec:supporting-diverse-programs}).
    \item \textbf{Pattern-Recognition-Based Compression:} We present a novel compression algorithm based on pattern recognition, capable of identifying and compressing recurring code patterns and common I/O patterns. The algorithm also includes an inter-process compression stage, effectively reducing redundancy across processes. With this new approach, Recorder achieves highly efficient compression, resulting in smaller trace sizes and improved scalability. (Section~\ref{sec:compression}).
    \item \textbf{Enabling In-depth Analyses:}  
    We explore how the detailed data collected by Recorder facilitates comprehensive I/O analyses. This includes potential analyses enabled by calls uniquely captured by Recorder, studies previously conducted using Recorder data, and two example case studies of workflow I/O behavior. (Section~\ref{sec:enabled-analyses}).
\end{itemize}

The rest of the paper is organized as follows. Section~\ref{sec:overview} provides an overall introduction to Recorder, covering its tracing mechanism, preserved information, and tracing features and capabilities. Section~\ref{sec:compression} discusses in detail the pattern-recognition-based compression algorithm. Section~\ref{sec:enabled-analyses} demonstrates how the detailed information collected by Recorder enables in-depth I/O studies. In Section~\ref{sec:evaluation}, we evaluate the scalability and overhead of Recorder using various I/O codes. Section~\ref{sec:related-work} discusses the related work. Finally, we conclude in Section~\ref{sec:conclusion}.

%% file: sections/overview.tex
\begin{figure}[htbp]
    \centering
    \includegraphics[width=0.8\linewidth]{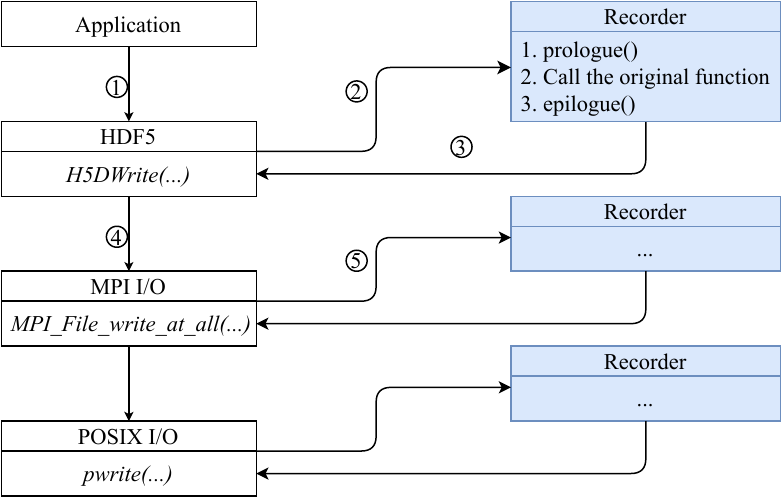}
    \caption{Example of instrumentation of the I/O stack by Recorder. \textcircled{1} Application calls the HDF5 library method \texttt{H5Dwrite}. \textcircled{2} Recorder intercepts the function and performs the tracing process. \textcircled{3} Recorder calls the real \texttt{H5Dwrite} function. \textcircled{4} \texttt{H5Dwrite} calls the MPI function \texttt{MPI\_File\_write\_at\_all}. \textcircled{5} \texttt{MPI\_File\_write\_at\_all} is also intercepted and recorded by Recorder. This continues until the I/O stack reaches the POSIX layer.}
    \label{fig:recorder_overview}
\end{figure}

 is implemented as a shared library that can be preloaded using LD\_PRELOAD on Linux systems, allowing it to be integrated into applications at runtime without requiring any code modifications or recompilation.
Figure~\ref{fig:recorder_overview} shows an example of the tracing process. In this example, the application makes an HDF5 call, which in turn invokes MPI-IO and POSIX calls, all of which are intercepted and stored by Recorder. 
Currently,  supports tracing for NetCDF, PnetCDF, HDF5, MPI-IO, MPI, and POSIX functions, with easy extensibility to include additional I/O layers, as we will show later. The tracing of each layer in Recorder can be dynamically enabled/disabled at runtime using environment variables.

\subsection{Automatic Tracing Code Generation}
\label{subsec:tracing-code-generation}
Recorder defines a tracing wrapper for each function to be intercepted. The actual interception is achieved using the GOTCHA library~\cite{GOTCHA}. When a function is intercepted, GOTCHA triggers the corresponding tracing wrapper (as shown by the blue boxes in Fig.~\ref{fig:recorder_overview}) for that function. 

The pseudocode for the tracing wrapper is shown in Listing~\ref{lst:tracing_wrapper}.
The wrapper is implemented as a macro, which removes dependencies on specific function signatures, making it easier to generate automatically.  The \texttt{prologue} (line 2) and \texttt{epilogue} (line 4) handle the storage of arguments before and after the function call, while the original function is invoked and its return value retrieved in between (line 5).

Both \texttt{prologue} and \texttt{epilogue} are implemented using macros to accommodate arbitrary argument types. Their implementations are identical for all intercepted functions, so they are defined only once. 
Specially, in \texttt{prologue}, we capture the function name, input parameters, entry time, etc. In \texttt{prologue}, we retrieve the exit time and store the return value. For Fortran MPI calls, we also need to set their error codes here, as the C wrappers have different function signatures compared to their Fortran counterparts (the latter include an additional output parameter for the error code). Finally, we create a \emph{record} containing all the collected information and apply an \emph{intra-process compression} to it. The details of the compression algorithm will be discussed in Section~\ref{sec:compression}.

\begin{minipage}{0.95\linewidth}
\vspace{1em}
\begin{lstlisting}[language=C,caption=Recorder tracing wrapper,label=lst:tracing_wrapper]
wrapper(func, ret_type, n_args, args) {
    prologue();
    ret_type ret = func(args);
    epilogue(n_args, args);
    return ret;
}
\end{lstlisting}
\end{minipage}

Writing a tracing wrapper manually for each targeted function is tiresome and error-prone, especially when covering large API sets (e.g., PnetCDF has over 900 APIs) or when extending support to additional I/O libraries. 
To address this, we developed a code-generation tool that takes a \textit{function signature file} (similar to a header file) as input and automatically generates a tracing wrapper for each function in the file. The generated code can be easily compiled as plugins, allowing Recorder to support new functions seamlessly. Table~\ref{tab:recorder-expansion} compares the number of functions from three supported I/O libraries in the current version of Recorder with those in its predecessor.

\begin{table}[htbp]
    \centering
    \caption{Supported functions of three I/O libraries. Recorder provides full coverage for HDF5, NetCDF, and PnetCDF APIs.}
    \label{tab:recorder-expansion}
    \begin{tabular}{|l|c|c|c|}
    \hline
    & HDF5 & NetCDF & PnetCDF \\
    \hline
    Recorder & 749 & 300 & 915 \\
    \hline
    Recorder-old & 84 & 0 & 0 \\
    \hline
    \end{tabular}
\end{table}

\subsubsection{Runtime Filtering}

As mentioned earlier, Recorder allows users to specify the I/O layers they want to intercept through environment variables, set before the targeted application's execution. During execution, Recorder also provides a filtering feature that gives users fine-grained control over which calls to trace and which to skip.

Users can define a list of file path prefixes they wish to monitor. Within the \texttt{prologue} function, Recorder performs a prefix match for function calls (e.g., \texttt{open} and \texttt{lstat}) that involve a file path parameter. A call is recorded if its file path parameter matches one of the user-specified prefixes. This feature is particularly useful for focusing on specific I/O operations, such as those involving large checkpoint files, while ignoring others. It also reduces tracing overhead and makes the resulting traces easier to analyze.

One challenge is that many I/O calls operate on file handles rather than file paths, requiring us to track the file handles associated with filtered paths. To address this, Recorder applies special handling to `open' calls (e.g., \texttt{fopen} and \texttt{MPI\_File\_open}) that are recorded, storing the generated file handles (such as \texttt{FILE*} and \texttt{MPI\_File}) in a hash set. Subsequently, when calls that reference a file handle (e.g., \texttt{fwrite} and \texttt{MPI\_File\_write\_at}) are intercepted, Recorder checks the hash set to determine whether to record the call.

\subsection{Supporting Diverse Programs}
\label{subsec:supporting-diverse-programs}

Recorder was initially designed to support single-threaded MPI applications, which represent a significant portion of HPC applications. In this revision, we have expanded its capabilities to include multi-threaded and non-MPI programs, broadening its applicability.

\textit{\textbf{Multi-threading Support:}}
Many HPC applications are multi-threaded, and while it is uncommon, some use different threads to perform asynchronous I/O. We carefully implemented all potentially concurrent operations, including interception and compression, in a thread-safe manner using locks. Additionally, for each intercepted call, we store the thread ID (of type \texttt{pthread\_t}), allowing post-processing tools to distinguish I/O calls from different threads and analyze the overlap of asynchronous I/O activities.

\textit{\textbf{Non-MPI Support:}}
Some HPC workflows, such as Montage~\cite{Montage_parallel}, include single-process, non-MPI programs where communication between programs is managed by workflow managers like Pegasus~\cite{Pegasus}. For Recorder to function properly, it needs to be initialized before tracing begins and finalized after the application's execution to clean up resources and write out trace files. In the case of MPI programs, Recorder automatically intercepts \texttt{MPI\_Init} (or \texttt{MPI\_Init\_thread}) and \texttt{MPI\_Finalize} to handle initialization and finalization. For non-MPI programs, Recorder employs the GNU C \emph{constructor} attribute to initialize before entering the \texttt{main()} function and the \emph{destructor} attribute to finalize after \texttt{main()} completes

\textit{\textbf{CUDA Support:}}
Applications often use GPUs to accelerate computation, while I/O operations are managed by the CPUs. When tracing GPU-enabled applications, capturing additional information about GPU kernels can provide valuable insights into the overlap between computation and I/O. Recorder supports intercepting CUDA kernels through the CUDA Profiling Tools Interface (CUPTI)~\cite{cupti}, treating kernel invocations as ordinary I/O calls. Currently, Recorder stores the kernel name (excluding parameters), as well as the entry and exit times. CUDA kernel tracing is optional and can be enabled or disabled at runtime.

\subsubsection{Extensive Metadata Information}
\label{subsubsec:preserved_information}
Recorder captures function names, parameters, and entry and exit timestamps for each intercepted call, along with extensive metadata. This metadata includes application-level details such as username, hostname, application name, and process ID, as well as record-level information like call depth and thread ID. Most of this information is straightforward to retrieve and compress. Here, we focus on two fields that require additional efforts to store.

\textit{\textbf{Call Depth:}}
Applications that use high-level I/O libraries may also make direct low-level I/O calls. For example, a POSIX I/O call can originate directly from the application, or be invoked by MPI-IO or HDF5. Tracking the call depth for each I/O operation provides insight into the function call flow through the I/O stack, highlighting cause-and-effect relationships across layers and helping to identify the source of performance bottlenecks. Recorder uses a stack-based approach to track call depth: when a new I/O call is intercepted, it is pushed onto the stack, incrementing the call depth; upon completion, the call is popped from the stack, decrementing the depth.

\textit{\textbf{Timestamps:}}
Recorder records the entry and exit times of each intercepted call, storing them as deltas relative to the application's start time. Each timestamp requires four bytes of memory and is stored in an internal buffer during the application's execution. At finalization, these timestamps are merged and compressed across all processes using zlib~\cite{zlib} and written to a single file using collective MPI-IO operations.

\subsection{Trace Conversions}
\label{subsec:post-processing-support}

The primary goal of an I/O tracing tool is to produce the application's I/O information with low overhead. This dictates the trace format, the fields stored, and the data types of the fields. However, these choices affect the cost of analyzing the trace data. In particular, the CFG- and CST-based traces (will be discussed in the next section) are good for compression and storage but can be less convenient for data analysis.

To make post-processing easier and also be able to take advantage of existing post-processing tools, we developed two trace format converters: the first is a Chrome timeline~\cite{Chrome_timeline} converter, and the second is a Parquet~\cite{Parquet} converter. Both are included within the Recorder package. The Chrome timeline is an event viewer tool that can visualize and query various events in the system. Chrome timeline uses a standard JSON format. Converting to this format also allows users to visualize Recorder traces in other Chrome timeline tools such as perfetto~\cite{Perfetto}. The Parquet converter converts the Recorder format into the Apache Parquet format, which is a column-oriented data file format. This format is great for doing large-scale analysis in the cloud~\cite{SeamlessRey2023}. In this converter, we decompress the Recorder trace files, retrieve each trace record, convert it into a parquet row, and write it to a parquet dataset. The converter creates a group of 64K records to generate ~100MB of files. Additionally, we set appropriate data types for each field within a record row. Finally, we apply Snappy compression to the logs to reduce the data size on the disk.

%% file: sections/compression.tex
As the number of iterations and the number of nodes on which an application runs increase, so does the number of function calls it makes. Without effective compression, the trace size will often grow linearly with the total computation amount. Fortunately, many codes exhibit some recurring patterns that, if recognized, can significantly reduce redundancies in the trace file. More importantly, when the same pattern is detected and compressed across processes, it is sometimes possible to achieve a constant trace size regardless of the execution time and scale. This is often the case for parallel I/O codes.

At runtime, Recorder intercepts each targeted call, rerouting it to an internal tracing routine where local (intra-process) pattern recognition and compression occur. At the finalization point, Recorder performs a global (inter-process) pattern recognition and compression pass, further reducing trace redundancies across processes.

The pattern recognition of Recorder can be divided into two categories: \emph{recurring pattern recognition} and \emph{I/O pattern recognition}. The former, adapted from~\cite{pilgrim_sc} and~\cite{ pilgrim_tpds}, detects and compresses recurring calling patterns such as loop structures. The latter identifies patterns specific to the I/O calls (e.g., the offsets of a sequence of \texttt{pwrite} calls). Recurring pattern recognition is performed locally, while I/O pattern recognition is performed both locally and globally.

\subsection{Recurring Pattern Recognition}

We introduce the recurring pattern recognition algorithm using a serial I/O example shown in Listing~\ref{lst:serial_io_pattern_example}. Running this example generates $m \times n$ identical \texttt{write} calls and $m$ identical \texttt{fsync} calls. Instead of storing each of these calls individually, Recorder recognizes the nested loop structure from the recurring calling pattern.

\begin{minipage}{0.95\linewidth}
\begin{lstlisting}[language=C,caption=A serial I/O example with nested loops,label=lst:serial_io_pattern_example]
for(int i = 0; i < m; i++) {
    for(int j = 0; j < n; j++)
        write(fd, buf, size);
    fsync(fd);
}
\end{lstlisting}
\end{minipage}

Recorder achieves this using a per-process \emph{context-free grammar} (CFG) and \emph{call signature table} (CST). A CFG is a formal grammar with production rules of the form $A \rightarrow \alpha$, where $A$ is a single non-terminal symbol, and $\alpha$ is a string of terminal and/or non-terminal symbols. Within the CFG, each terminal symbol represents a unique \emph{call signature}, which consists of record-level information including function name, function parameters, thread ID, and call depth. The CST maintains the association between each terminal symbol and its corresponding call signature.

Essentially, building each local CFG and CST identifies and compresses the process local recurring patterns. Table~\ref{tab:serial_io_cfg_cst} shows the generated CFG and CST for the code in Listing~\ref{lst:serial_io_pattern_example}. The CST contains two entries, each representing a unique call signature. Note that in the CST here and later, we only show function names and function parameters, omitting call depth and thread ID for simplicity. The CFG has two rules: $S$, representing the outer loop, and $A$, representing the inner loop. $S$ is also called the \emph{starting rule}, from which, by performing recursive rule applications, one can recover the original sequence of I/O calls.

\begin{table}[htbp]
    \small
    \centering
    \caption{Generated CFG and CST of Listing~\ref{lst:serial_io_pattern_example}.}
    \label{tab:serial_io_cfg_cst}
    \begin{tabular}{|l|l|}
        \hline
        CFG & CST \\ \hline
        $S \rightarrow A^m$ & $a$: \texttt{write(fd, buf, size)} \\
        $A \rightarrow a^n b $ & $b$: \texttt{fsync(fd)} \\ \hline
    \end{tabular} 
\end{table}

Recorder employs the well-known Sequitur algorithm~\cite{Sequitur} to construct a CFG that describes the string of I/O calls. The Sequitur algorithm is an online algorithm that processes one terminal symbol (i.e., one call) at a time, exhibiting linear time complexity in terms of the number of symbols. For the CST, we implemented it using a hash table with call signatures as keys and terminal symbols as values.
At execution time, when a new call is intercepted, we gather all necessary information, construct its call signature, and consult the local CST to obtain its corresponding terminal symbol. If the call signature already exists in the local CST, we retrieve its terminal symbol; otherwise, we create a new entry and assign it an unused terminal symbol. Finally, we append the terminal symbol to the local CFG using the Sequitur algorithm.

\subsection{I/O Pattern Recognition}

Two calls to the same I/O function with different parameters will be encoded by different terminals. But the parameters of such calls may obey a simple relation, providing opportunities for further compression. For example, Listing~\ref{lst:parallel_io_pattern_example} shows a common way of performing parallel I/O, where $nprocs$ processes write to a shared file in a strided manner with a fixed chunk size. The base offset is computed according to the rank of each process (line 1), and the stride size is determined based on the total number of processes (line 2). Before each \texttt{write} call, a \texttt{lseek} is invoked to reposition the file pointer of the calling process to the correct position.

\begin{minipage}{0.95\linewidth}
\begin{lstlisting}[language=C,caption=A parallel I/O example,label=lst:parallel_io_pattern_example]
size_t base = rank * chunk_size;
size_t stride = nprocs * chunk_size;
for(int i = 0; i < m; i++) {
    lseek(fd, base+stride*i, SEEK_SET);
    write(fd, buf, size);
}
\end{lstlisting}
\end{minipage}

\begin{figure}[htbp]
    \centering
    \includegraphics[width=0.85\linewidth]{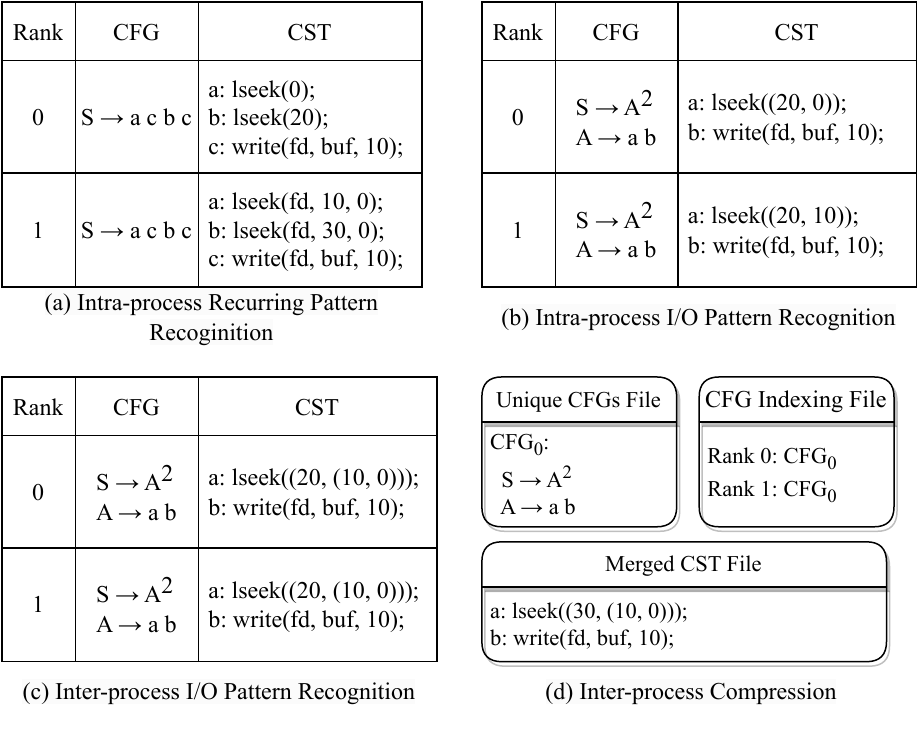}
    \caption{Recorder compression steps for Listing~\ref{lst:parallel_io_pattern_example}. The intra-process recurring pattern recognition (a) and the intra-process I/O pattern recognition (b) are executed at runtime, while the inter-process I/O pattern recognition (c) and the subsequent inter-process compression step (d) are performed during the finalization stage. For simplicity, the CSTs only display the offset parameter for \texttt{lseek} calls.}
    \label{fig:Recorder_compression_steps}
\end{figure}

Assuming we run this example with $nprocs = 2, m = 2$, and $chunk\_size = 10$, 
Figure~\ref{fig:Recorder_compression_steps}(a) displays the generated CFG and CST of each process. Each CST consists of three entries, one for each unique I/O call signature. The CFGs have only one rule that represents the sequence of calls. It is clear that the loop structure is not recognized due to the two \texttt{lseek} calls having different offsets.
In general, for this example, each CFG will have $2m$ symbols and each CST will have $m+1$ entries. Consequently, the trace size will grow linearly with both the number of processes and the number of iterations.

To address this issue, in addition to recurring pattern recognition, Recorder also performs I/O pattern recognition. The basic idea is to identify the offset patterns observed from the same I/O calls and encode them in a way that produces an \emph{identical} call signature. There are two types of I/O patterns we need to detect: intra-process I/O pattern and inter-process I/O pattern.

\subsubsection{Intra-process I/O Pattern}

This pattern recognizes I/O calls that differ only in their offset, where the offset of the $i$-th call follows the form $i\times a + b$, with $a$ and $b$ as constants for the given process. Since $a$ and $b$ have the same values for all calls of the same pattern, they can be easily calculated by examining two adjacent calls. 
Recorder performs an intra-process I/O pattern check whenever intercepting an I/O call containing an offset parameter. Recorder checks whether the current offset is equal to $i\times a + b$, assuming that this is the $i$-the call of the same pattern. If so, Recorder encodes the offset as a $(a, b)$ pair; otherwise, the original offset is stored. 
For example, the offsets of rank 0's \texttt{lseek} calls in Figure~\ref{fig:Recorder_compression_steps}(a) can be calculated as $i\times 20 + 0$. Thus we can encode their offsets as (20, 0) and remove the duplicated call signatures. With the intra-process pattern recognition, the local CFGs can recognize the loop structure (Figure~\ref{fig:Recorder_compression_steps}(b)). As a result, the sizes of each CFG and CST will stay unchanged regardless of the number of iterations.

\subsubsection{Inter-process I/O Pattern}
To facilitate inter-process compression, which will be discussed in the next subsection, it is necessary to identify I/O patterns across processes. Such patterns are common in parallel I/O codes, where distinct processes access a shared file at different locations. In these scenarios, the offsets of the same I/O call made by different processes often follow linear functions of the caller's rank, taking the form of $rank \times a + b$.
Inter-process I/O pattern recognition involves communication between processes and is not executed at the time of I/O interception. Recorder performs this check at the finalization point. If the offset is already encoded in the form of $(a, b)$ during the intra-process I/O pattern recognition step, then the inter-process I/O pattern check will be performed separately on $a$ and $b$. Figure~\ref{fig:Recorder_compression_steps}(c) illustrates the resulting CFGs and CSTs with inter-process I/O pattern recognition.

Finally, a notable challenge arises due to the presence of opaque MPI file handles. Consider two ranks executing \texttt{MPI\_File\_write\_at()} to write to a shared file. The file is specified by an \texttt{MPI\_File} type parameter, which was collectively opened using \texttt{MPI\_File\_open()}. Assuming that the offset patterns are recognizable across the two ranks and all other parameters are identical, a complication still arises with the \texttt{MPI\_File} handle.
This handle is unique to the local process, but distinct ranks may possess different \texttt{MPI\_File} handles that refer to the same file. Consequently, the inter-process pattern recognition will not be able to merge these two calls since the file handle is part of the call signature. To address this issue, during each \texttt{MPI\_File\_open()}, the process with rank 0 in that invoking group of processes determines a group-wide unique ID for the newly created file handle and broadcasts it to every other rank. Subsequently, all ranks in the same communicator associate this unique ID with their local file handle, utilizing a local hash table. For all subsequent MPI I/O calls that access this file handle, Recorder replaces it with its associated unique ID when composing the call signature.

\subsection{Inter-process Compression}

After application execution, each process retains its individual CFG and CST, which are often identical across processes owing to the pattern recognition technique described earlier. To eliminate redundancies across processes, a final inter-process compression is performed.

\subsubsection{Inter-process CST compression}
First, the root process (rank 0) gathers the CSTs from all other processes and consolidates them into a single CST according to the call signatures (i.e., the value of each CST entry). The call signatures may receive new terminal symbols as they are merged into the new CST. 
Next, the merged CST is distributed back to each process, allowing them to adjust their local CFG with the updated terminal symbols. Finally, the root process writes out the merged CST to a `merged CST file'.

\subsubsection{Inter-process CFG compression}
Following the update of each process's local CFG, Recorder conducts a uniqueness check to identify replicated CFGs and keeps only one copy of each CFG. A CFG index is saved, storing for each rank the index of the corresponding CFG.

In conclusion, the final inter-process compression phase produces three essential files, as illustrated in Figure~\ref{fig:Recorder_compression_steps}(d). For typical I/O scenarios, the size of the merged CST and the number and sizes of the unique CFGs remain constant irrespective of the number of processes and iterations. Recorder generates two more files: one for storing timestamps of all intercepted calls (Section~\ref{subsubsec:preserved_information}) and another for saving additional application-level and Recorder runtime information (e.g., the enabled tracing layers and the version of Recorder).

%% file: sections/enabled_analyses.tex
A natural question when developing yet another tracing tool is:
\emph{What new analyses does it enable that are not easily achievable with existing tools?}
We address this question from three perspectives: (1) discussing potential analyses made possible by supporting a broader range of function calls, (2) highlighting several studies conducted over the years using Recorder traces, and (3) presenting two case studies based on real-world HPC workflows.

\subsection{Potential Analyses Enabled by Additional Functions}
We use POSIX functions as an example to illustrate how support for additional function calls can enable deeper analyses. The comparison of supported functions is based on Recorder, Darshan, and Score-P, as detailed in~\cite{liem_mango-io_2023}.

Table~\ref{tab:recorder_posix_calls} lists the POSIX calls that are exclusively captured by Recorder, along with their potential analysis use cases.
These calls are organized into logical groups based on their functionality, such as directory management, file ownership and permissions, symbolic and hard links, access control checks, and memory synchronization. Capturing these calls allows for more in-depth analysis of key I/O behaviors in HPC applications.

For instance, examining directory management calls (e.g., \verb|mkdir|, \verb|rmdir|, \verb|chdir|) can reveal patterns related to the creation and traversal of directories during intermediate stages of scientific workflows. 
Paul et al.~\cite{paul_understanding_2020} demonstrated a strong correlation between metadata operations and writes, particularly for \verb|mkdir| and \verb|mknod|. 
Tracking file permission changes (e.g., \verb|chmod|, \verb|chown|) helps identify potential access control issues.
Similarly, capturing memory synchronization (\verb|msync|) allows for the identification of inefficiencies in out-of-core data processing. 
This level of granularity is particularly valuable for understanding I/O behavior and identifying potential performance bottlenecks in HPC applications.

\begin{table}[htbp]
\small
\centering
\caption{POSIX calls exclusively captured by Recorder}
\label{tab:recorder_posix_calls}
\begin{tabular}{|l|l|}
\hline
\textbf{POSIX Calls} & \textbf{Analysis Cases} \\ \hline
\begin{tabular}[c]{@{}l@{}}mkdir, rmdir, chdir, \\ opendir, readdir, rewinddir\end{tabular} & \begin{tabular}[c]{@{}l@{}}Tracking the creation and deletion of temporary \\ directories for intermediate results\end{tabular} \\ \hline
\begin{tabular}[c]{@{}l@{}}chmod, chown, \\ lchown\end{tabular} & \begin{tabular}[c]{@{}l@{}}Tracking changes in file permissions to ensure \\ proper data isolation\end{tabular} \\ \hline
\begin{tabular}[c]{@{}l@{}}link, linkat, symlink, \\ symlinkat, readlink, \\ readlinkat\end{tabular} & \begin{tabular}[c]{@{}l@{}}Monitoring the usage of symbolic links for \\ managing complex data and software \\ dependencies\end{tabular} \\ \hline
access, faccessat & \begin{tabular}[c]{@{}l@{}}Identifying potential bottlenecks caused by \\ frequent access checks\end{tabular} \\ \hline
msync & \begin{tabular}[c]{@{}l@{}}Analyzing memory-mapped file synchronization \\ in out-of-core data processing\end{tabular} \\ \hline
pipe, mkfifo & \begin{tabular}[c]{@{}l@{}}Analyzing producer-consumer patterns in \\ parallel processing pipelines\end{tabular} \\ \hline
tmpfile & \begin{tabular}[c]{@{}l@{}}Tracking temporary file usage patterns to \\ optimize burst buffer configurations\end{tabular} \\ \hline
truncate, ftruncate & \begin{tabular}[c]{@{}l@{}}Monitoring file size adjustments in storage \\ allocation\end{tabular} \\ \hline
umask, utime & \begin{tabular}[c]{@{}l@{}}Analyzing file creation patterns and timestamp \\ modifications\end{tabular} \\ \hline
mknod, mknodat & \begin{tabular}[c]{@{}l@{}}Analyzing the usage of device files for direct \\ I/O operations\end{tabular} \\ \hline
\end{tabular}
\end{table}


\subsection{Example Studies using Recorder Traces}

The first study~\cite{devarajan_extracting_2022} used the Dask parallel computing library~\cite{Dask} to perform efficient analysis of traces. The traces were first converted into Parquet format using Recorder's trace conversion tool (see Section~\ref{subsec:post-processing-support}).
The Parquet files were loaded in an out-of-core fashion and analyzed on a distributed cluster. 
Using Dask's DataFrame API, we ran Pandas-like queries on large-scale traces that could not fit in memory. 
We analyzed terabytes of Recorder traces, extracting I/O patterns such as file size distribution across two billion files, read-write distribution of the workload, and more. 
This approach enabled detailed data dependencies to be identified within the HPC workflow and pinpointed I/O bottlenecks from different perspectives---such as timelines, processes, and files. 
Enabled by Recorder's rich data, these perspectives allowed us to compute key metrics, including aggregate transfer size, bandwidth, and I/O time across the distributed workflow.

In IOMax~\cite{yildirim_iomax_2023}, building upon the analysis work above, we optimized I/O analysis performance using Recorder's comprehensive traces.
These detailed traces allowed us to implement query reduction strategies to minimize redundant I/O operations and employ caching mechanisms to accelerate repeated queries.
Additionally, we transformed raw Recorder traces into optimized formats, enabling faster data slicing and more efficient querying.
These improvements were crucial for handling terabytes of trace data, which traditional tools struggle to process at scale.
With these optimizations, IOMax achieved up to 7\texttimes{} improvement in out-of-core data drilling, demonstrating how Recorder's fine-grained trace data can be leveraged to unlock new insights and streamline large-scale I/O analysis.

The final example~\cite{yellapragada:2021:verifying-io,wang:2021:consistency-semantics} explores storage consistency semantics~\cite{wang:2024:formal}. During the development of a new parallel file system~\cite{brim:2023:unifyfs}, we conducted an analysis to understand the various I/O calls made by HPC applications and their associated consistency semantics requirements. 
This information is particularly valuable for file system designers, as it can inform implementation decisions and optimization strategies.
Existing tools fell short in this regard, either by omitting certain calls or failing to capture specific parameters needed for consistency requirements, such as the precise offset of each I/O call and \verb|flag|s/\verb|mode|s associated with \verb|open| calls.

\subsection{Case Studies of HPC Workflows}

\textit{1000 Genomes}: is a computationally intensive workflow that processes large datasets of human genomes to identify genetic variants and mutations, leveraging parallel processing capabilities to maximize analysis throughput~\cite{the_1000_genomes_project_consortium_map_2010}.
For this workflow, Recorder captured 646M POSIX calls issued across 21M files by 2.7K processes. 
Notably, 109M (17\%) of these calls are metadata calls.
Of those metadata calls, a subset of 6.5M (6\%) were only detectable by Recorder, with the most frequent being \verb|access|, \verb|opendir|, \verb|mkdir|, \verb|unlink|, \verb|ftell|, \verb|fcntl|, \verb|rmdir|, and \verb|chmod| calls.
Our analysis reveals that 99\% of these \verb|chmod| calls are triggered by the \textit{individuals\_merge} application during the final step of writing merged chromosome files with the necessary permissions (i.e. \verb|chmod 420|), following the extraction of compressed genomic data archives and the creation of temporary directories required for the process.
Additionally, our analysis shows that the \textit{individuals} and \textit{individuals\_merge} applications issue \verb|fcntl| calls with the \verb|F_DUPFD_CLOEXEC| argument on file descriptors for temporary directories to manage inheritance during process execution before removing them.

\textit{Montage}: is a mosaic engine that converts sky-survey data into PNG images, using a 6-stage workflow with parallel processing~\cite{jacob_montage_2009}.
The Montage workflow consists of both MPI and non-MPI  applications, which are both supported by Recorder to provide a comprehensive data flow.
Specially, for the workflow we ran, Recorder captured 6.4M POSIX calls issued across 10K files by 10 cooperating applications.
The application excessively issues small read and write requests (\textless{}4KB).
Notably, 258K (4\%) of these POSIX calls are metadata calls.
Among the metadata calls analyzed, approximately 27K (10\%) were only detectable by Recorder, with the most frequent being \verb|pipe|, \verb|unlink|, and \verb|remove| operations.
Our analysis shows that only a specific subset of applications, namely \textit{mAddMPI}, \textit{mDiffExecMPI}, \textit{mFitExecMPI}, and \textit{mProjExecMPI}, make use of \verb|pipe| calls.
These calls are implemented to facilitate efficient inter-process communication (IPC), optimizing data transfer between processes.
This unconventional use of \verb|pipe| alongside MPI suggests a hybrid communication strategy within Montage, potentially aimed at reducing overhead and enhancing workflow efficiency.
This level of insight would not have been achievable without Recorder's detailed tracing capabilities, which capture \verb|pipe| calls missed by other tools~\cite{liem_mango-io_2023}.

%% file: sections/evaluation.tex
In comparison to existing tools, the additional capabilities and comprehensive information preserved by Recorder enable more thorough I/O analyses. As previously mentioned, we have conducted numerous I/O studies over the years using Recorder traces, with detailed analyses outlined in those studies. In this section, our main focus is on evaluating the proposed compression algorithm and the scalability of Recorder.

We evaluate Recorder using a parallel I/O benchmark, IOR~\cite{ior}, and a real-world scientific application, FLASH~\cite{app_flash}. We first examine the proposed I/O pattern recognition algorithm using IOR (Section~\ref{subsec:impact_of_io_pattern}). Next, we evaluate the pattern-recognition-based compression as a whole using FLASH simulations at large scales (Section~\ref{subsec:scalability}). Finally, we compare Recorder with Recorder-old and Darshan from the perspective of both space and time overhead (Section~\ref{subsec:comparison}).

In all experiments, Recorder intercepts and store I/O calls from HDF5, MPI-IO, and POSIX layers. The trace records preserve the values of all parameters of the intercepted calls. Compared to Recorder-old, Recorder stores additional record-level and application-level information as discussed in Section~\ref{subsubsec:preserved_information}. For Darshan runs, we enable the STDIO, H5F, H5D, DXT-POSIX, and DXT-MPIIO modules, which cover a similar set of functions as intercepted by Recorder. It is worth mentioning that Recorder do not distinguish STDIO calls from POSIX calls; they are all considered POSIX calls.
With the DXT mode enabled, Darshan is able to capture finer-grained details (e.g., exact offset and number of bytes accessed per call) but not every parameter of the call. Also, the DXT mode currently only supports POSIX and MPI-IO calls, and it focuses mainly on data operations. In other words, Darshan in DXT mode still collects less information than Recorder. While the comparison with Darshan is not apples-to-apples, it provides a good baseline for evaluating Recorder's overhead.

The experiments in Section~\ref{subsec:impact_of_io_pattern} and Section~\ref{subsec:comparison} were conducted on Quartz at Lawrence Livermore National Laboratory. Quartz is a Penguin system consisting of 2988 computing nodes. Each node has 2 18-core Intel Xeon E5-2695 processors (2.1 GHz) and 128 GB of memory.
The experiments in Section~\ref{subsec:scalability} were conducted on Theta at Argonne National Laboratory. Theta is a Cray XC40 system consisting of 4,392 Intel KNL 7230 compute nodes. Each compute node has 64 cores and 192 GB of memory.
We used the same version of Darshan (v3.4.3) and Recorder-old (v2.2.1) on both systems. All I/O operations in our experiments were supported by the Lustre parallel file system.

\subsection{Impact of I/O Pattern Recognition}
\label{subsec:impact_of_io_pattern}

To assess the effectiveness of the I/O pattern recognition algorithm, the trace size reported in this subsection includes only the `unique CFGs file' and the `merged CST file.' These two files store all function calls, their parameters, call depths, thread IDs, etc. We exclude the `CFG indexing file' and the `timestamp file' as they are expected to grow with the process count and their sizes are not affected by the pattern recognition compression.

First, we use IOR to study the impact of the I/O pattern recognition algorithm on trace size. Here, all processes used POSIX I/O to write to a single shared file, with varying block sizes and process counts. The recurring pattern recognition and the final inter-process compression step are always performed.

\begin{figure}[htbp]
    \centering
    \begin{subfigure}{0.45\linewidth}
        \includegraphics[width=\linewidth]{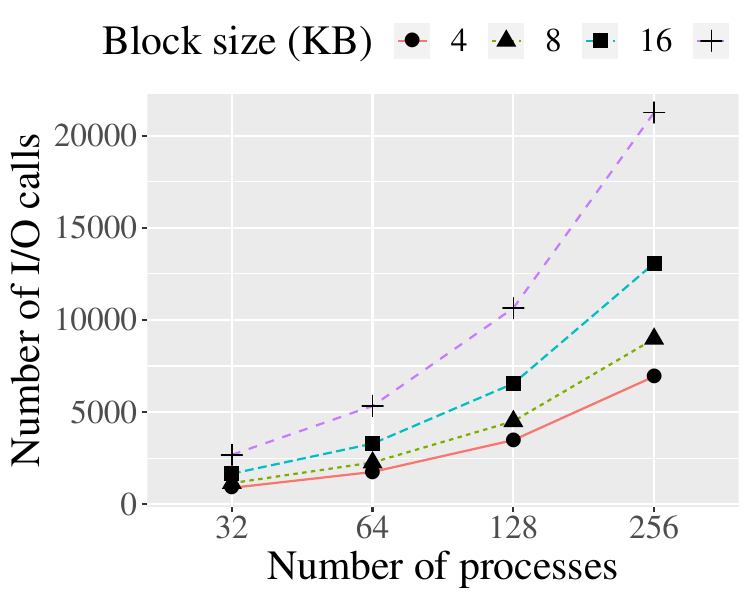}
    \end{subfigure}%
    \begin{subfigure}{0.45\linewidth}
        \includegraphics[width=\linewidth]{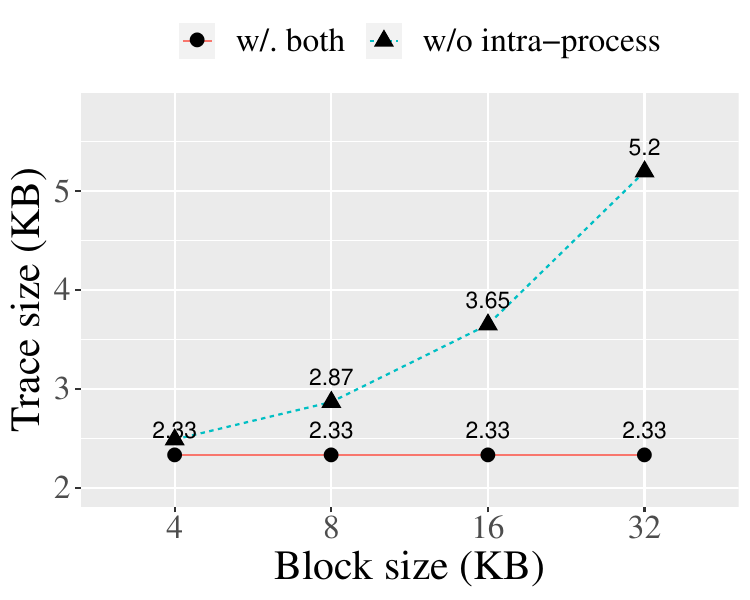}
    \end{subfigure}
    \caption{Impact of the intra-process I/O pattern recognition on trace size. The left figure depicts the number of intercepted calls for various block sizes and process counts. On the right, with the process count fixed at 256, the trace sizes are displayed across different block sizes. With inter-process I/O pattern recognition only, the trace size increased with block size due to a higher number of generated I/O calls.}
    \label{fig:ior_trace_size_vs_block_size}
\end{figure}

Figure~\ref{fig:ior_trace_size_vs_block_size} shows the impact of intra-process I/O pattern recognition on trace size. As observed on the left, since the transfer size was fixed, the number of I/O calls made by IOR increased linearly with the block size, resulting in an increase in trace size. However, with intra-process I/O pattern recognition, the trace size remained unchanged as all the increased I/O calls were identified and compressed using the same I/O pattern.

\begin{figure}[htbp]
    \centering
    \includegraphics[width=0.9\linewidth]{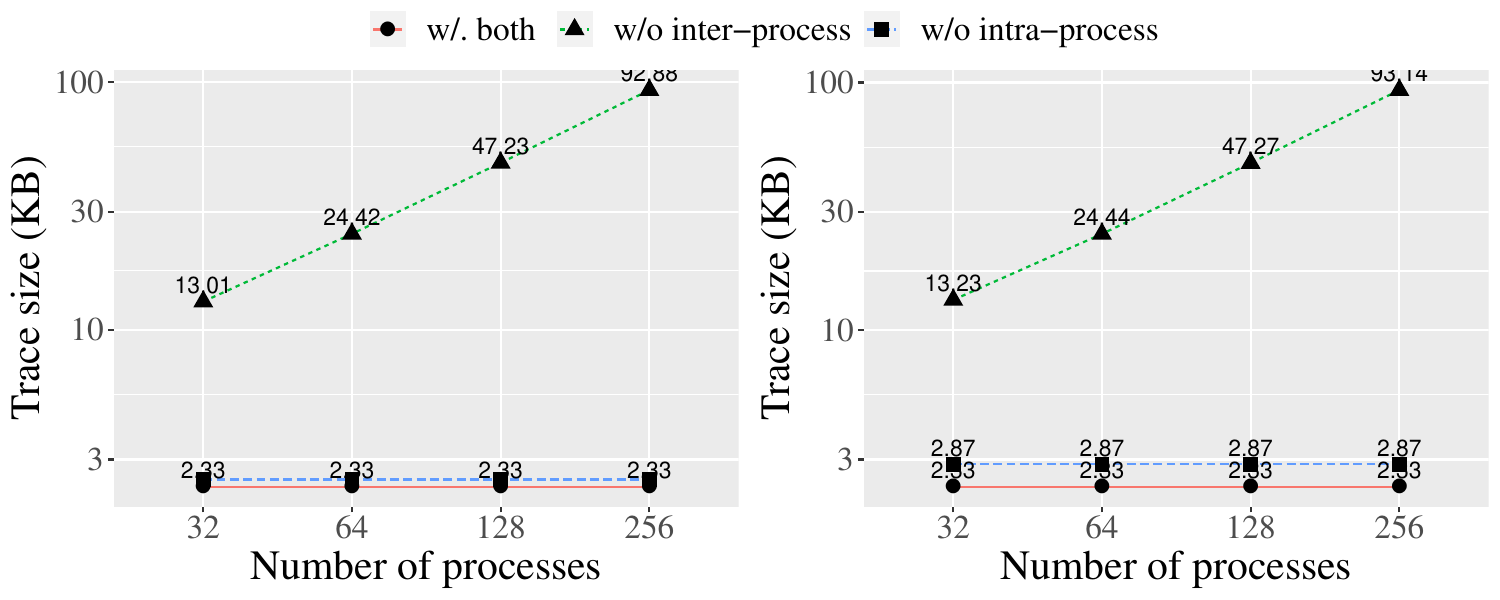}
    \caption{Impact of inter-process I/O pattern recognition. The figures show trends for 4KB (left) and 8KB (right) block sizes. With a fixed block size, intra-process I/O pattern recognition has no impact on the scalability, while inter-process I/O pattern recognition ensures a constant trace size regardless of the process count.}
    \label{fig:ior_trace_size_vs_procs}
\end{figure}

Figure~\ref{fig:ior_trace_size_vs_procs} shows the trace sizes of three different Recorder configurations with an increasing number of processes. For a fixed block size, the trace size grew linearly when inter-process I/O pattern recognition was disabled. However, disabling only the intra-process I/O pattern recognition resulted in a slightly larger but constant trace size. With both I/O pattern recognition enabled, the trace size was at its minimum and unchanged regardless of the number of processes.

\subsection{Scalability of Recorder}
\label{subsec:scalability}

The previous subsection demonstrates the effectiveness of I/O pattern recognition on compressing traces of simple HPC I/O codes. In this subsection, we enable all pattern recognition techniques and put Recorder to the test with a real-world application (FLASH) running at large scales. Again, the trace size here is the sum of the final CFG and CST sizes.

We use two FLASH simulations: Cellular and Sedov; both are shipped with the FLASH package and have been configured to run on 3D domains with weak scaling.
Each run of a FLASH simulation outputs several files, including log files, plot files, and checkpoint files. In our experiments, we focus on the I/O calls that write plot files and checkpoint files. The other files are extremely small and take negligible time to write.

FLASH uses HDF5 to write the plot files and checkpoint files, which in turn issues MPI-IO calls, which then invoke POSIX I/O calls to perform the actual I/O operations.
FLASH supports two I/O modes: (1) Independent I/O mode, which uses independent MPI-IO calls such as \texttt{MPI\_File\_write\_at}. (2) Collective I/O mode, which uses collective MPI-IO calls such as \texttt{MPI\_File\_write\_at\_all}. The two modes exhibit significantly different I/O patterns, which present different challenges to the compression algorithm. We evaluate Recorder with both modes to show the effectiveness of our compression algorithm.

\subsubsection{Independent I/O}

\begin{figure}[htbp]
    \centering
    \begin{subfigure}{0.45\linewidth}
        \includegraphics[width=\linewidth]{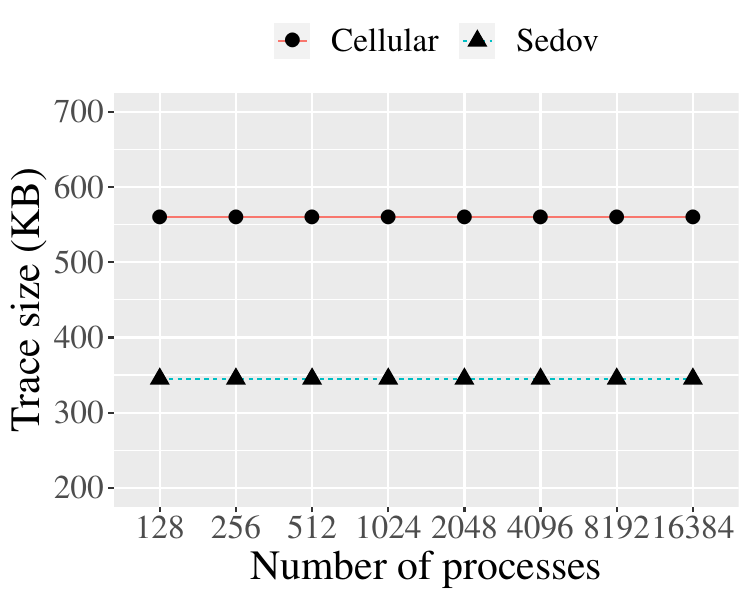}
    \end{subfigure}%
    \begin{subfigure}{0.45\linewidth}
        \includegraphics[width=\linewidth]{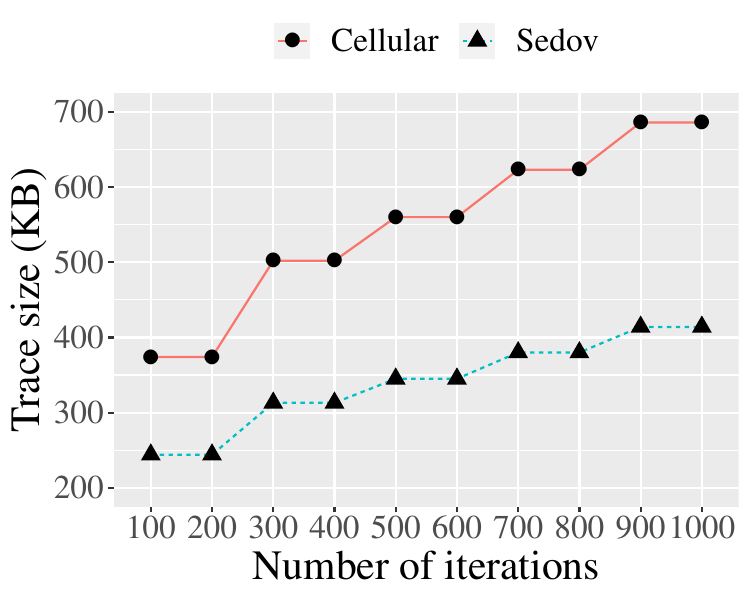}
    \end{subfigure}
    \caption{Evaluation of Recorder's I/O trace size with FLASH simulations. Independent MPI-I/O was used for all runs. When varying the process count (left), the number of iterations was fixed at 500. When varying the number of iterations (right), the process count was fixed at 16K.}
    \label{fig:exp_flash_io_trace_size}
\end{figure}

In this mode, HDF5 issues independent MPI-IO calls in which each process writes its own portion of data to a shared output file (plot file or checkpoint file). The problem size per process was fixed (weak scaling), so both the total number of I/O calls and the final output file size increased linearly in the number of processes.
Figure~\ref{fig:exp_flash_io_trace_size} shows how Recorder's trace size scales with the number of processes and the number of iterations.
On the left, we kept the number of iterations fixed at 500 and increased the number of processes from 128 to 16,384. Although the number of intercepted I/O calls increased linearly in the number of processes, the trace file size remained constant. The reason is that the same I/O patterns were present at all scales in this weak scaling experiment. Each process performed identical writes but to different file offsets, which resulted in a linear pattern that was easily recognized and encoded using the writer's rank.
On the right, we ran the simulations on 16K processes and increased the number of iterations from 100 to 1000. The simulations were configured to output a new plot file and checkpoint file every 200 iterations. Since every new file uses a different name, and the filename is also included in the call signature, Recorder saw a new set of I/O call signatures every 200 iterations. We can clearly see that the Recorder's I/O trace size jumped at every 200$\times$ iterations. The increase in trace size in this case can easily be avoided. One way is to write output files in a rolling manner.
For example, we can use the ``rolling\_checkpoint'' option of FLASH to keep a maximum of two checkpoints at a time, where an older checkpoint is overwritten by a new one.
This way, the trace file size will stop increasing after the second checkpointing step. However, this method does not work when all output files are to be preserved, such as snapshots and plot files. In such cases, we can leave the filename out of the call signature and store it separately. Alternatively, we could identify and compress the pattern in the file names, which is often a linear pattern (e.g., ``plot-$i$'' for the $i$-th plot file). Either way, the I/O call signatures of different files become identical, thus writing new files will not result in new I/O patterns.

\subsubsection{Collective I/O}
In this mode, HDF5 issues collective MPI-IO calls, where only a subset of processes are selected to perform the actual I/O operations. Those processes are called aggregators. Each aggregator first collects the data from a \emph{distinct} set of processes, including itself. Then it groups the data according to the offset and performs writes in larger chunks. The collective I/O increases the I/O chunk size and reduces the number of POSIX I/O operations. 

The aggregator count is an important parameter for I/O performance tuning. Running with a different number of aggregators will generate different I/O patterns. The MPI-IO implementation in our system, ROMIO, uses the Lustre stripe count and the number of compute nodes to decide the aggregator count. The Lustre stripe count is a user-tunable parameter that controls how many Lustre data servers are used to store a file. At runtime, ROMIO calculates the aggregator count using the minimum of the stripe count and the number of compute nodes for this run.

\begin{figure}[tbp]
    \centering
    \includegraphics[width=0.9\linewidth]{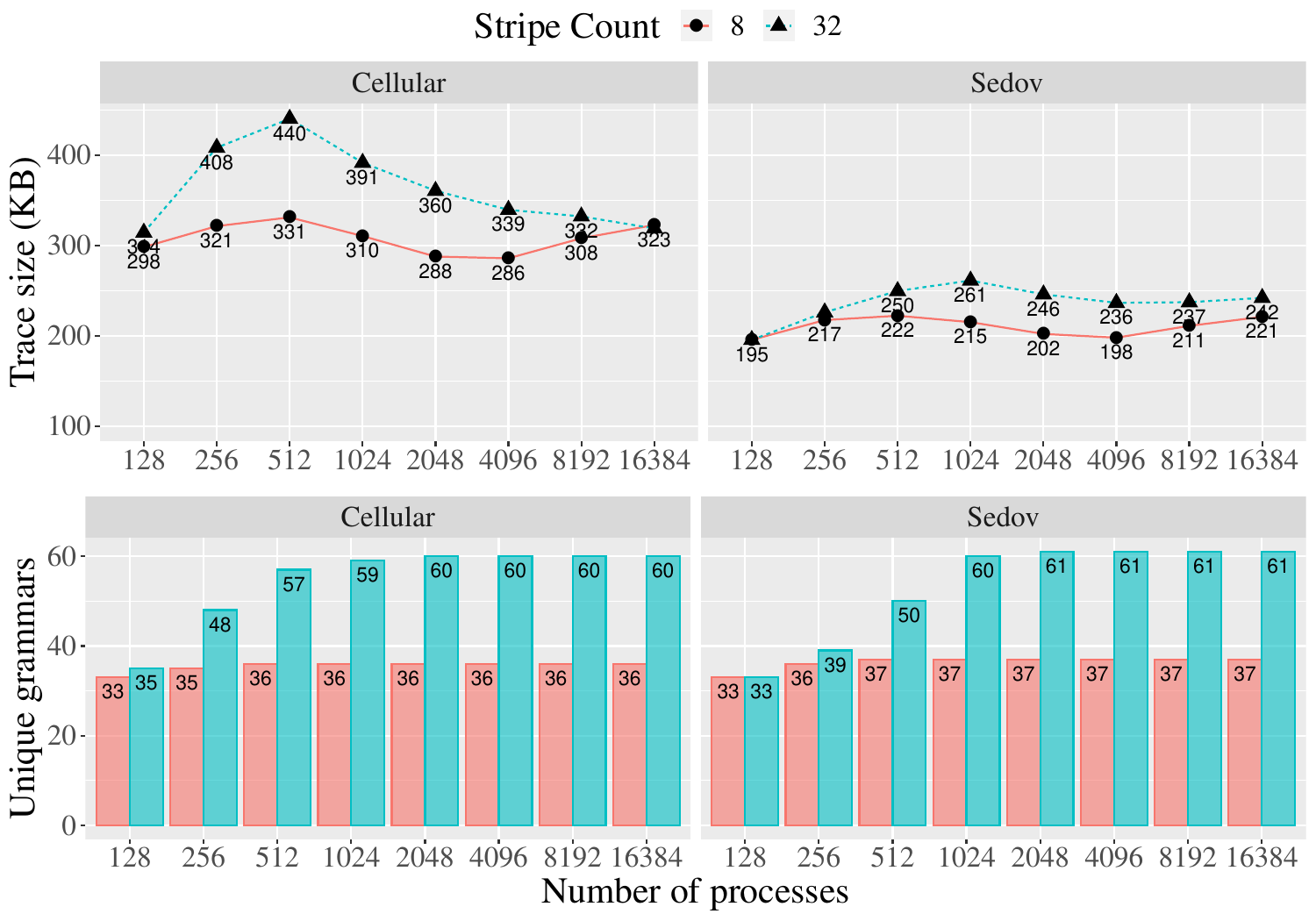}
    \caption{Evaluation of Recorder's I/O trace size with FLASH simulations. Collective MPI-I/O was enabled for all runs. Two different stripping configurations were used: red color for stripe count 8 and blue color for stripe count 32. The top figures show how the trace size scales with the number of processes. The bottom figures show the number of unique grammars stored in the merged CFG file.}
    \label{fig:exp_flash_io_trace_size_vs_procs}
\end{figure}

In our experiments, we evaluate Recorder using two stripe counts: 8 and 32.
Figure~\ref{fig:exp_flash_io_trace_size_vs_procs} shows the results. The top two figures show how the trace size of each FLASH simulation changed with the number of processes. Unlike the independent I/O experiment, where the trace size stayed constant, in the collective I/O experiments, we observe that in both simulations, the trace size fluctuated with small rises and falls.
The number of aggregators increases with the number of nodes until it reaches the stripe count, either 8 or 32. The number of I/O patterns depends on the number of aggregators and does not increase once their number stops increasing, at 8 nodes (512 processors) for stripe count 8 and 32 nodes (2,048 processes) for stripe count 32.
This is confirmed by the bottom figures. They show that in both simulations, the number of unique grammars (CFGs) stopped increasing after 512 and 2,048 processes for the stripe count of 8 and 32.

\begin{figure}[htbp]
    \centering
    \includegraphics[width=0.95\linewidth]{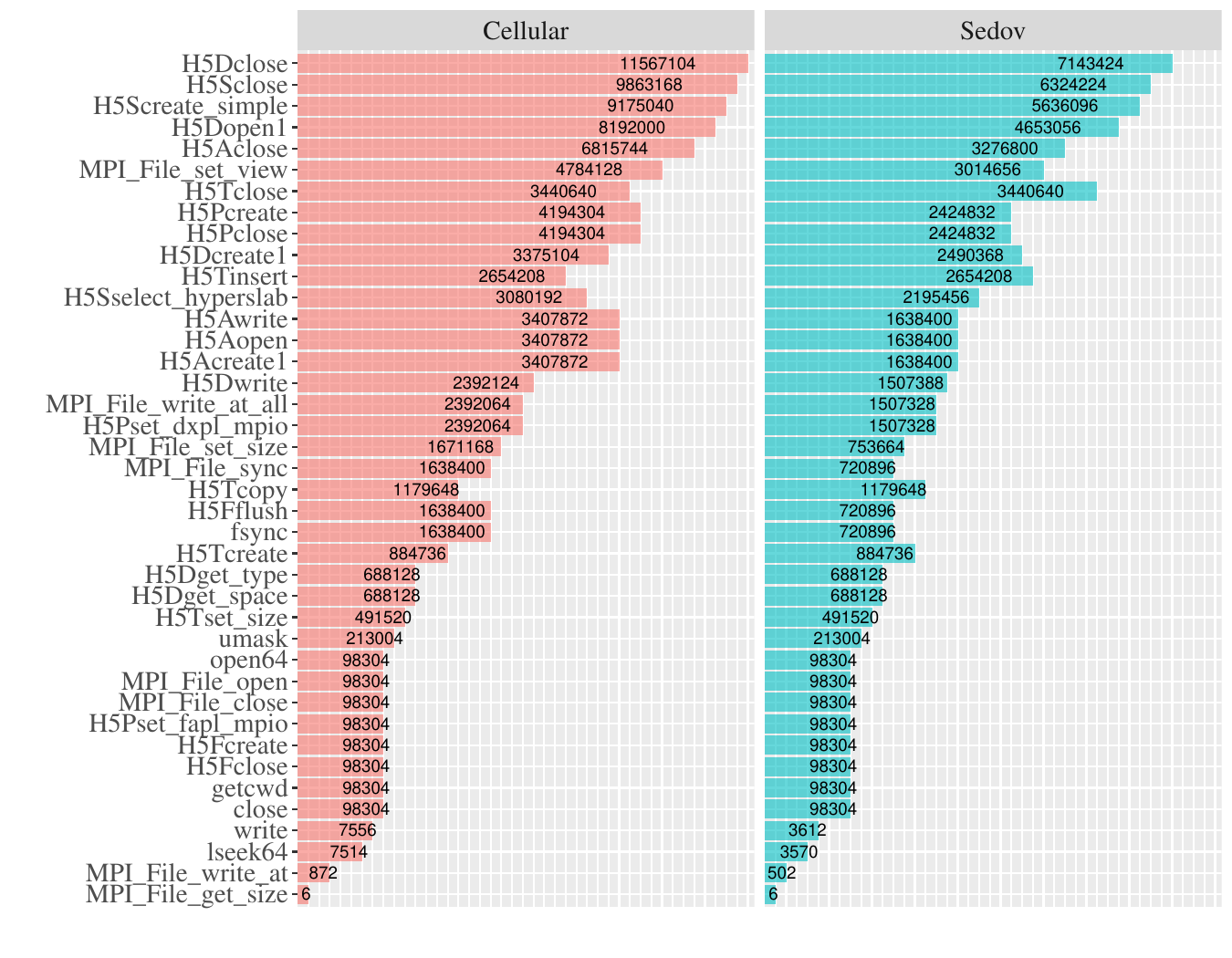}
    \caption{I/O calls count of 16K-process FLASH simulations. Both simulations use an identical set of I/O functions.}
    \label{fig:exp_flash_io_calls_count}
\end{figure}

\begin{figure}[htbp]
    \centering
    \includegraphics[width=0.9\linewidth]{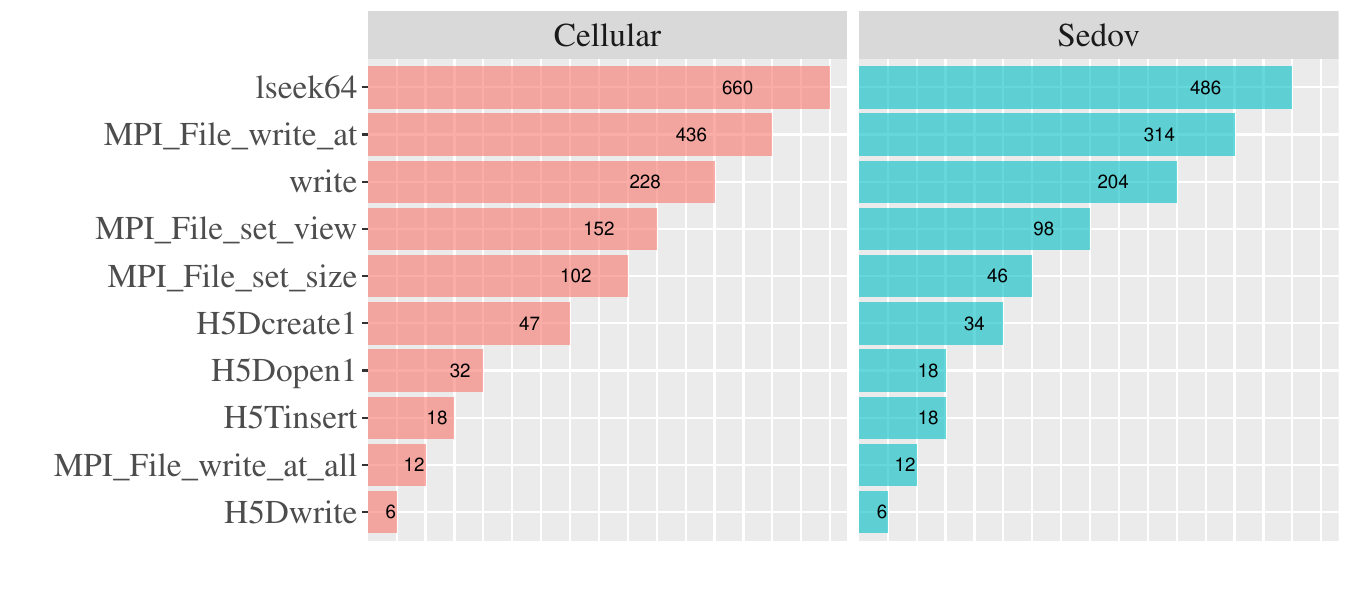}
    \caption{Top ten I/O functions with the highest call signature count. Cellular made more calls and had more unique call signatures thus producing a larger CST file.}
    \label{fig:exp_flash_io_calls_signature}
\end{figure}

As mentioned earlier, the reported trace size here is the sum of the CFG file size and the CST file size. The CFG file size depends on the number of unique grammars (and of course the total number of grammar rules and symbols). As we can see, the two simulations produced very different trace sizes even with a similar number of unique grammars. The discrepancy was caused by their CST files. The two simulations used an identical set of I/O calls (due to the programming modularity of FLASH) as shown in Figure~\ref{fig:exp_flash_io_calls_count}. But the Cellular simulation made many more I/O calls than Sedov. However, a higher call count does not necessarily make a larger CST file, as frequent calls may be easy to compress. The CST file size is decided by the number of unique call signatures. 
We show in Figure~\ref{fig:exp_flash_io_calls_signature} the top ten I/O functions that are responsible for the most unique call signatures. For example, \texttt{H5Dclose} was the most frequent function in both simulations, but \texttt{lseek64} had the highest number of unique call signatures, making it the most difficult function to compress. Nevertheless, Cellular again produced more unique call signatures and thus had larger total trace sizes.

\subsection{Comparison with Recorder-old and Darshan}
\label{subsec:comparison}

Finally, we compare Recorder with Recorder-old and Darshan in terms of space and time overhead. Regarding space overhead, our focus is on comparing the size of traces produced by the three tools for the same application. As for time overhead, we compare the total execution time of application runs with and without the three tools.

\subsubsection{Space Overhead}
Here, the reported trace size for Recorder includes every file produced, that is, not only the CFG and CST files but also the timestamps and CFG indexing files. The size of the timestamps file is linear in the number of calls and is dominant for large runs. We conducted the comparison using the same two FLASH simulations (Cellular and Sedov). Both simulations ran for 100 iterations, generating a checkpoint and plot file every 20 iterations. As before, we present results for both collective I/O mode and independent I/O mode.

Table~\ref{tab:trace_size_comparison} shows the resulting trace sizes of Cellular and Sedov. Consistent with the analysis in the previous subsection, Cellular made more I/O calls than Sedov, resulting in larger trace files for all three tools. Additionally, for each tool, the trace size roughly doubles as we double the number of processes. This increase is attributed to the storage of timestamps, which grows linearly (two timestamps per I/O call) with the process count.

\input{tables/trace_size_comparison}

When compared to Recorder-old, Recorder stored more information and yet achieved a significantly smaller trace size in every run, thanks to the pattern-recognition-based compression. On average, Recorder's trace size was $12\times$ smaller than that of Recorder-old.
Compared to Darshan, in collective I/O mode, the trace size of Recorder (for both simulations) was about $6$ to $7\times$ larger than Darshan's traces, which is expected since Recorder captures much more information than Darshan, including more HDF5 calls and all function parameters. Interestingly, when switched to independent I/O mode, Darshan's trace size increased significantly, while Recorder's trace size stayed relatively stable, resulting in a difference of only about $3\times$. Examination of Darshan traces using the `darshan-parser' tool revealed that the most significant increase in trace size came from the DXT\_POSIX module. For instance, in a 64-process Sedov run, the trace generated by the DXT\_POSIX module occupies 21,038 bytes in collective mode, whereas it takes 115,317 bytes in independent mode.

\subsubsection{Time Overhead}
The time overhead is attributed to the additional time introduced by the profiling or tracing tool, calculated as a percentage relative to the execution time of an application without any tool. Typically, the overhead arises at three points: (1) initialization time, (2) finalization time, and (3) each time an I/O call is intercepted and stored. The initialization and finalization phases may involve global communications, such as broadcasting configurations and merging files, but they occur only once per application run and can be amortized as the application execution time increases. In contrast, runtime tracing overhead occurs every time an I/O call is intercepted, with its overall impact heavily dependent on the ratio of I/O to communication and computation. Applications with higher I/O operation counts are expected to experience more significant runtime overhead.

\begin{figure}[tbp]
    \centering
    \includegraphics[width=0.7\linewidth]{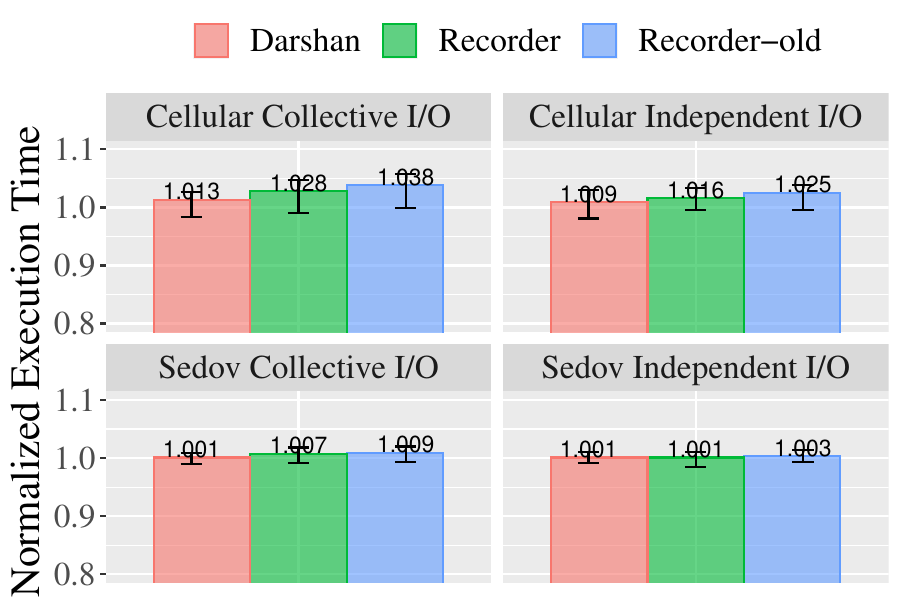}
    \caption{Time overhead comparison with FLASH simulations.}
    \label{fig:flash_time_overhead_comparison}
\end{figure}

To compare the time overhead of the three tools, we conducted multiple runs of FLASH simulations, each time with a different tool or with no tool. Each run was repeated at least ten times.
The runs without any tool had a relatively constant execution time with increasing scale when using weak scaling.
The observed overhead of all three tools also remained stable. Figure~\ref{fig:flash_time_overhead_comparison} shows the normalized execution time (execution time with tools divided by execution time without tools) of FLASH simulations running with 4096 processes. Cellular simulations took approximately twice as long to finish as Sedov and also exhibited a slightly larger variance. On average, Darshan showed the smallest overhead, with Recorder falling between Darshan and Recorder-old. While Recorder's overhead is slightly higher than Darshan, the maximum observed overhead was only about $3\%$. It's worth noting that we intentionally performed very frequent checkpointing (5 times in a short period), significantly increasing the execution time ratio between I/O and other components. In reality, with less frequent checkpoints, the overhead would be even lower, as the I/O time (thus runtime tracing overhead) would constitute a smaller proportion of the overall execution time.

%% file: tables/trace_size_comparison.tex
\begin{table}[tb]
    \centering
    \small
    \caption{Comparison of the trace sizes of Recorder, Recorder-old, and Darshan using Cellular and Sedov simulations.}
    \label{tab:trace_size_comparison}
    \begin{tabular}{l|ccc|ccc}
      \toprule
      \multicolumn{1}{c}{} & \multicolumn{6}{c}{\textbf{Cellular}} \\
      \multicolumn{1}{c}{} & \multicolumn{3}{c}{Collective I/O} & \multicolumn{3}{c}{Independent I/O} \\
      \midrule
      Procs & Darshan & Recorder & Recorder-old & Darshan & Recorder & Recorder-old \\
      64   & 248K & 1.5M &	18M  & 404K & 1.5M & 18M  \\ 
      128  & 416K &	2.8M &	36M  & 738K & 2.8M & 35M  \\	
      256  & 753K &	5.5M &	71M  & 1.4M & 5.3M & 68M  \\
      512  & 1.4M &	11M  &	141M & 2.7M	& 11M  & 136M \\ 
      1024 & 2.7M &	22M  &	282M & 5.2M	& 21M  & 272M \\	
      2048 & 5.8M &	43M  &	565M & 11M	& 41M  & 546M \\
      4096 & 12M  & 90M  &	1.2G & 21M  & 83M  & 1.1G \\ 
      \bottomrule
      \toprule
      \multicolumn{1}{c}{} & \multicolumn{6}{c}{\textbf{Sedov}}    \\
      \multicolumn{1}{c}{} & \multicolumn{3}{c}{Collective I/O} & \multicolumn{3}{c}{Independent I/O} \\
      \midrule
      
      Procs & Darshan & Recorder & Recorder-old & Darshan & Recorder & Recorder-old \\ 
      64   & 150K  & 940K & 12M  & 244K & 930K & 11M  \\
      128  & 250K  & 1.8M & 23M  & 449K & 1.8M & 22M  \\
      256  & 481K  & 3.4M & 46M  & 898K & 4.0M & 44M  \\
      512  & 907K  & 6.7M & 91M  & 1.8M & 6.5M & 87M  \\
      1024 & 1.8M  & 14M  & 181M & 3.5M & 13M  & 173M \\
      2048 & 3.9M  & 27M  & 366M & 6.9M & 26M  & 347M \\
      4096 & 8.9M  & 57M  & 738M & 14M  & 52M  & 694M \\
      \bottomrule
    \end{tabular}
\end{table}

%% file: sections/related_work.tex
There have been significant research efforts in the area of I/O tracing and profiling tools. On one hand, system-level tools such as iotop~\cite{iotop} and iostat~\cite{iostat} monitor system-level I/O activities and report metrics across the entire system. On the other hand, application-level tools, including Darshan~\cite{Darshan}, Score-P~\cite{Score-p}, IPM-IO~\cite{uselton2010ipmio}, and IOPin~\cite{kim2012iopin}, run alongside an application and collect information about that specific application. Here, we specifically focus on application-level tools as they are more closely related to our work.

Darshan~\cite{Darshan} and TAU~\cite{TAU} are profiling tools that record both POSIX I/O and MPI-IO activities, providing statistics for individual applications. Darshan goes a step further by supporting additional layers, including HDF5, Lustre, etc. Although these profiling tools collect statistics with low overhead and offer a good estimate of overall performance, they may lack the detailed information required for in-depth analysis, such as function parameters and entry/exit times for each call. Darshan addresses this limitation by developing two DXT modules (DXT-POSIX and DXT-MPIIO) to collect more detailed information. However, these DXT modules do not cover all layers nor preserve all necessary function parameters.

IPM-I/O~\cite{uselton2010ipmio} extends the capabilities of the existing performance tool IPM~\cite{IPM} by incorporating I/O operation tracing. While IPM-I/O traces POSIX I/O calls, it necessitates applications to be linked against the IPM-I/O library.
VampirTrace~\cite{VampirTrace} records I/O functions of the standard C library and also supports tracing GPU-accelerated applications. ScalaIOTrace~\cite{ScalaIOTrace} supports both MPI-IO and POSIX I/O, offering the added advantage of generating compressed event logs. 
Score-P~\cite{Score-p} is another popular tool suite for profiling, event tracing, and online analysis of HPC applications. It works with other tools such as TAU~\cite{TAU}, Vampir, and Scalasca~\cite{Scalasca}.
Several of the aforementioned tools, including Score-P, Vampir, and TAU, support a tracing format named OTF~\cite{otf1} or its optimized versions such as OTF2~\cite{otf2}, OTFX~\cite{otfx}, and enhanced OTF2~\cite{enhanced_otf2}. While OTF employs zlib compression to reduce trace size, it is a general-purpose compression algorithm and lacks inter-process compression. Some other work~\cite{mohror2009trace_compression, weber2013alignment} proposed to compress traces by detecting the similarities of different sections of traces. However, tools relying on such techniques generally fall short in providing structure-aware and I/O-aware compression, leading to reduced compression rates.
In contrast to the above tools, Recorder intercepts NetCDF, PnetCDF, HDF5, MPI, and POSIX I/O calls without requiring modification or recompilation of applications. Compared to its predecessor, Recorder captures more functions and stores extensive application-level and record-level metadata information. More importantly, Recorder employs a sophisticated pattern-recognition-based algorithm to compress traces effectively for recurring I/O patterns.

%% file: sections/conclusion.tex
In this work, we presented Recorder, a comprehensive parallel I/O tracing and analysis tool. Recorder supports multi-threaded codes and works for both MPI and non-MPI programs, aiming to enable a comprehensive I/O study for HPC workflows. Recorder employs a new pattern-recognition-based compression algorithm specifically designed for HPC I/O. We discussed our efforts for post-processing support, including additional trace information and two trace format converters. 
Through a thorough evaluation of our compression algorithm, we showcased its effectiveness in handling typical I/O patterns. Notably, Recorder was demonstrated to store more information than its predecessor while occupying significantly less storage space. Furthermore, our experiments showed that, as a tracing tool, Recorder incurs a comparable overhead when compared to a proofing tool.

Evaluations of some features, such as CUDA kernel inceptions, non-MPI program tracing and post-processing support, are left as future work. We also plan to evaluate Recorder with applications that may not benefit from the pattern-recognition-based compression, such as machine learning applications. For those applications, Recorder may experience higher overhead.

%% file: sections/acknowledgment.tex
This work was performed under the auspices of the U.S. Department of Energy by Lawrence Livermore National Laboratory under Contract DE-AC52-07NA27344 and was supported by the LLNL-LDRD Program under Project No. 23-ERD-053.
LLNL-JRNL-858585.
This work was supported by NSF SHF Collaborative grant 1763540.
This material is based upon work supported by the U.S. Department of Energy, Office of Science, Office of Advanced Scientific Computing Research under the DOE Early Career Research Program. This research used resources of the Argonne Leadership Computing Facility, which is a DOE Office of Science User Facility supported under Contract DE-AC02-06CH11357.